\documentclass[english,handout,notheorems]{beamer}
\usepackage{pgfpages}
\pgfpagesuselayout{2 on 1}[a4paper,border shrink=5mm]
\usepackage{amsmath}
\usepackage{amsthm}
\usepackage{amssymb}
\usepackage{bm}
\usepackage{xspace}
\usepackage{xcolor}
\usepackage{graphicx}
\usepackage{url}
\usepackage{framed}
\usepackage{float}
\usepackage{rotating}
\usepackage{verbatim}
\usepackage{listings}
\usepackage{lscape}
\usepackage{enumerate}

\theoremstyle{plain}
\newtheorem{proposition}{Proposition}
\DeclareMathOperator{\expop}{\mathbb{E}}
\DeclareMathOperator{\entop}{\mathbb{H}}

\DeclareMathOperator{\kl}{D}
\DeclareMathOperator{\probop}{Pr}

\newcommand{\executeiffilenewer}[3]{%
\ifnum\pdfstrcmp{\pdffilemoddate{#1}}%
{\pdffilemoddate{#2}}>0%
{\immediate\write18{#3}}\fi%
}

\graphicspath{{figures/}}

\definecolor{bblue}{rgb}{0.2,0.2,0.7}

\title[]{Rooted Trees with Probabilities Revisited}
\author[]{Georg B\"ocherer}
\institute[]{Institute for Communications Engineering\\Technische Universit\"at M\"unchen\\Email: \url{georg.boecherer@tum.de}}

\DeclareMathOperator{\supp}{supp}

\definecolor{darkblue}{RGB}{23,111,193}

\begin{document}

\begin{frame}
	\titlepage
\end{frame}

\begin{frame}
\frametitle{Abstract}
Rooted trees with probabilities are convenient to represent a class of random processes with memory. They allow to describe and analyze variable length codes for data compression and distribution matching. In this work, the Leaf-Average Node-Sum Interchange Theorem (LANSIT) and the well-known applications to path length and leaf entropy are re-stated. The LANSIT is then applied to informational divergence. Next, the differential LANSIT is derived, which allows to write normalized functionals of leaf distributions as an average of functionals of branching distributions. Joint distributions of random variables and the corresponding conditional distributions are special cases of leaf distributions and branching distributions. Using the differential LANSIT, Pinsker's inequality is formulated for rooted trees with probabilities, with an application to the approximation of product distributions. In particular, it is shown that if the normalized informational divergence of a distribution and a product distribution approaches zero, then the entropy rate approaches the entropy rate of the product distribution.
\end{frame}

\begin{frame}
\frametitle{Probability notation}
\begin{itemize}
\item Random variable $X$, takes values in $\mathcal{X}$
\item Distribution $P_X$: for each $a\in\mathcal{X}\colon P_X(a):=\probop(X=a)$.
\item Support $\supp P_X:=\{a\in\mathcal{X}\colon P_X(a)>0\}$.
\end{itemize}
\end{frame}

\begin{frame}
\frametitle{Rooted Trees with Probabilities \cite{rueppel1994leaf,masseyapplied1,kramer2012information}}
\begin{itemize}
\item $\mathcal{L}$: set of leaves.
\item $L$: random variable over $\mathcal{L}$.
\item We identify $\supp P_L\triangleq \mathcal{L}$, i.e., a node is a leaf of a tree if it has no successors and is generated with non-zero probability. 
\item $\mathcal{N}$: set of all nodes on paths through the tree.
\item root $0\in\mathcal{N}$.
\item $\mathcal{B}=\mathcal{N}\setminus\mathcal{L}$: set of branching nodes.
\item $\mathcal{L}_j$: leaves below node $j\in\mathcal{N}$. $j\in\mathcal{L}\Rightarrow\mathcal{L}_j=j$.
\item $\mathcal{S}_j$: successors of node $j\in\mathcal{B}$.
\item $S_j$: random variable over successors of node $j\in\mathcal{B}$. We identify $\mathcal{S}_j\triangleq \supp P_{S_j}$.
\end{itemize}
\end{frame}

\begin{frame}
\frametitle{Node Probabilities}
\begin{itemize}
\item We associate with each node $j\in\mathcal{N}$ a probability 
\begin{align}
Q_j = \sum_{i\in\mathcal{L}_i}P_L(i)
\end{align}
\item The probabilities of the successors of node $j\in\mathcal{B}$ are given by
\begin{align}
Q_i = Q_jP_{S_j}(i),\quad i\in\mathcal{S}_j.\label{eq:qp}
\end{align}
\end{itemize}
\end{frame}

\begin{frame}
\frametitle{Example}
\begin{figure}
\footnotesize
\centering
\def\svgwidth{0.8\textwidth}
\executeiffilenewer{figures/trees.svg}{figures/trees.pdf}%
{inkscape -z -D --file=figures/trees.svg %
--export-pdf=figures/trees.pdf --export-latex}%
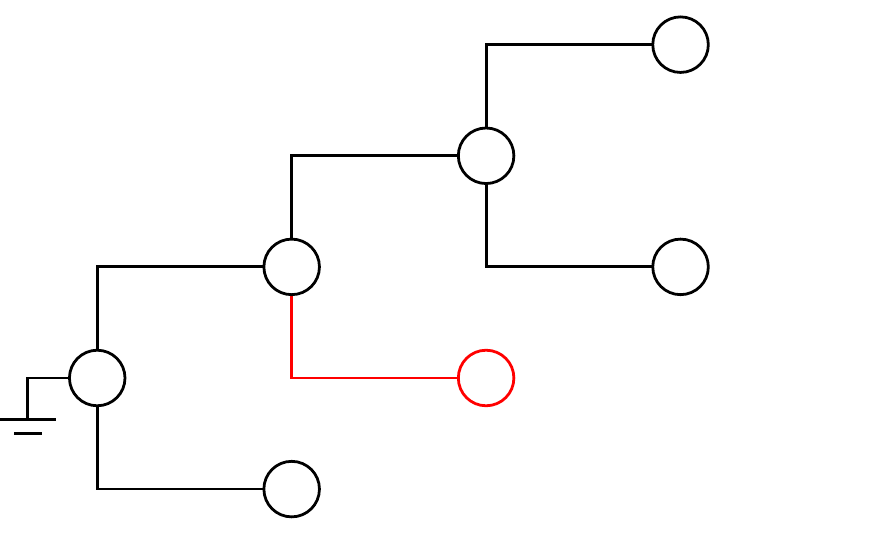%

\end{figure}
\parbox{0.44\textwidth}{
\begin{itemize}
\item $\supp P_L=\mathcal{L}=\{2,5,6\}$
\item $\mathcal{N}=\{0,1,2,3,5,6\}$
\item $\mathcal{B}=\mathcal{N}\setminus\mathcal{L}=\{0,1,3\}$
\item $\supp P_{S_1}=\mathcal{S}_1=\{3\}$
\end{itemize}
}
\hfill
\parbox{0.52\textwidth}{
\begin{itemize}
\item $\mathcal{L}_1=\{5,6\}$
\item $Q_1={\displaystyle\sum_{i\in\mathcal{L}_1}P_L(i)}=\frac{1}{2}+\frac{1}{4}=\frac{3}{4}$
\item $Q_3=Q_1P_{S_1}(3)=\frac{3}{4}\cdot 1=\frac{3}{4}$
\end{itemize}
}
\end{frame}

\begin{frame}[plain]
\begin{center}
\color{bblue}{\Large Leaf-Average Node-Sum Interchange Theorem\\
(LANSIT)}
\end{center}
\end{frame}

\begin{frame}
\frametitle{LANSIT \cite[Theorem 1]{rueppel1994leaf}}
\begin{itemize}
\item Let $f$ be a function that assigns to each node $j\in\mathcal{N}$ a real value $f(j)$.
\item For each $j\in\mathcal{N}\setminus 0$, define $\Delta f(j):=f(j)-f(\text{predecessor of } j)$
\end{itemize}
\begin{proposition}[LANSIT]\label{prop:integrallansit}
\begin{align}
\expop[f(L)]-f(0)=\sum_{j\in\mathcal{B}}Q_j\expop[\Delta f(S_j)]
\end{align}
\end{proposition}
\end{frame}

\begin{frame}
\frametitle{Proof of LANSIT}
\begin{itemize}
\item Consider a tree with leaves $\mathcal{L}$.
\item  Let $\mathcal{S}_j\subseteq\mathcal{L}$ be a set of leaves with a common predecessor $j$. 
\begin{align}
\sum_{i\in\mathcal{S}_j}P_L(i)f(i)\overset{(a)}{=}&\sum_{i\in\mathcal{S}_j}Q_jP_{S_j}(i)[f(i)-f(j)+f(j)]\\
=&Q_jf(j)\underbrace{\sum_{i\in\mathcal{S}_j}P_{S_j}(i)}_{=1}+Q_j\sum_{i\in\mathcal{S}_j}P_{S_j}(i)\Delta f(i)\\
=&Q_jf(j)+Q_j\expop[\Delta f(S_j)]
\end{align}
where (a) follows from \eqref{eq:qp}.
\item $\mathcal{L}\leftarrow j\cup\mathcal{L}\setminus\mathcal{S}_j$ is a new tree with a reduced number of leaves and $P_L(j)=Q_j$.
\item Repeat the procedure until $j$ is the root node $0$. Then $Q_j=1$ and $Q_jf(j)=f(0)$. 
\end{itemize}
\hfill$\square$
\end{frame}

\begin{frame}
\frametitle{Path Length Lemma \cite[Lemma 2.1]{kramer2012information}}
\begin{itemize}
\item Function $w(j):=$ length of path to node $j$.
\item For each $j\in\mathcal{N}\setminus 0$: $\Delta w(j)=1$.
\item $w(0)=0$.
\end{itemize}
\begin{proposition}[Path Length Lemma]\label{prop:integralpll}
\begin{align}
\expop[w(L)]=\sum_{j\in\mathcal{B}}Q_j.
\end{align}
\end{proposition}
\end{frame}

\begin{frame}
\frametitle{Leaf Entropy Lemma \cite[Lemma 2.2]{kramer2012information}}
Function $f(i)=-\log_2 Q_i$.
\begin{proposition}[Leaf Entropy Lemma]\label{prop:lel}
\begin{align}
\entop(P_L)=\sum_{j\in\mathcal{B}}Q_j\entop(P_{S_j}).
\end{align}
\end{proposition}
\small
\textcolor{bblue}{Proof.}
\begin{align}
\entop(P_L)=\expop[-\log_2 P_L(L)]=\expop[f(L)]&\overset{(a)}{=}\sum_{j\in\mathcal{B}}Q_j\expop[\Delta f(S_j)]\\
&=\sum_{j\in\mathcal{B}}Q_j\expop[-\log_2\frac{Q_{S_j}}{Q_j}]\\
&\overset{(b)}{=}\sum_{j\in\mathcal{B}}Q_j\expop[-\log_2 P_{S_j}(S_j)]\\
&=\sum_{j\in\mathcal{B}}Q_j\entop(P_{S_j})
\end{align}
where (a) follows by the LANSIT and (b) by \eqref{eq:qp}.\hfill$\square$
\end{frame}

\begin{frame}
\frametitle{Informational Divergence}
Function $f(i)=\log_2 \frac{Q_i}{Q'_i}$.
\begin{proposition}\label{prop:ldl}
\begin{align}
\kl(P_L\Vert P_{L'})=\sum_{j\in\mathcal{B}}Q_j\kl(P_{S_j}\Vert P_{S'_j}).
\end{align}
\end{proposition}
\footnotesize
\textcolor{bblue}{Proof.}
\begin{align}
\kl(P_L\Vert P_{L'})=\expop\Bigl[\log_2\frac{P_L(L)}{P_{L'}(L)}\Bigr]&\overset{(a)}{=}\sum_{j\in\mathcal{B}}Q_j\expop[\Delta f(S_j)]\\
&=\sum_{j\in\mathcal{B}}Q_j\expop[\log_2\frac{Q_{S_j}}{Q_j}\frac{Q'_j}{Q'_{S_j}}]\\
&\overset{(b)}{=}\sum_{j\in\mathcal{B}}Q_j\expop[\log_2\frac{P_{S_j}(S_j)}{P_{S'_j}(S_j)}]\\
&=\sum_{j\in\mathcal{B}}Q_j\kl(P_{S_j}\Vert P_{S'_j})
\end{align}
where (a) follows from the LANSIT and (b) by \eqref{eq:qp}.\hfill$\square$
\end{frame}

\begin{frame}
\frametitle{Random Vectors}
\label{slide:remark}
\textcolor{bblue}{Remark.} If all paths in a tree have the same length $n$,  then $P_L$ can be thought of as a joint distribution $P_{X^n}$ of a random vector $X^n=(X_1,\dotsc,X_n)$. In this case, Prop.~\ref{prop:lel} and Prop.~\ref{prop:ldl} are the chain rules for entropy and informational divergence, respectively.
\end{frame}

\begin{frame}[plain]
\begin{center}
\color{bblue}{\Large Differential LANSIT}
\end{center}
\end{frame}

\begin{frame}
\frametitle{Differential LANSIT}
\begin{itemize}
\item $B$: random variable over branching nodes $\mathcal{B}$.
\item Define
\begin{align}
P_B(j)=\frac{Q_j}{\expop[w(L)]},\quad j\in\mathcal{B}.
\end{align}
\item By path length lemma
\begin{align}
\sum_{j\in\mathcal{B}}P_B(j)=\frac{\sum_{j\in\mathcal{B}}Q_j}{\expop[w(L)]}=1.
\end{align}
\item[$\Rightarrow$] $P_B$ defines a distribution over $\mathcal{B}$.
\end{itemize}
\begin{proposition}[Differential LANSIT]
\begin{align}
\frac{\expop[f(L)]-f(0)}{\expop[w(L)]}=\expop[\Delta f(S_B)].
\end{align}
\end{proposition}
Note that the expectation on the right-hand side is over $P_{S_B}P_B$.
\end{frame}

\begin{frame}
\frametitle{Example}
\begin{itemize}
\item Consider the path length function $w$. 
\item By the Differential LANSIT,
\begin{align}
\expop[\Delta w(S_B)]=\frac{\expop[w(L)]-w(0)}{\expop[w(L)]}=\frac{\expop[w(L)]}{\expop[w(L)]}=1.
\end{align}
\end{itemize}
\end{frame}

\begin{frame}
\frametitle{Entropy Rate}
Function $f(i)=-\log_2 Q_i$.
\begin{proposition}\label{prop:dlel}
\begin{align}
\frac{\entop(P_L)}{\expop[w(L)]}&=\expop[\entop(P_{S_B})]
\end{align}
\end{proposition}
\textcolor{bblue}{Proof.}
\begin{align}
\frac{\entop(P_L)}{\expop[w(L)]}=\frac{\expop[-\log_2 P_L(L)]}{{\expop[w(L)]}}\overset{(a)}{=}&\expop[-\log_2\frac{Q_{S_B}}{Q_B}]\\
=&\expop[-\log_2\frac{Q_{S_B}}{Q_B}]\\
\overset{(b)}{=}&\expop[-\log_2 P_{S_B}(S_B)]\\
=&\expop[\entop(P_{S_B})]
\end{align}
where (a) follows by the differential LANSIT and (b) by \eqref{eq:qp}.\hfill$\square$
\end{frame}

\begin{frame}
\frametitle{Normalized Informational Divergence}
Function $f(i)=\log_2 \frac{Q_i}{Q'_i}$.
\begin{proposition}\label{prop:ndivergence}
\begin{align}
\frac{\kl(P_L\Vert P_{L'})}{\expop[w(L)]}&=\expop[\kl(P_{S_B}\Vert P_{S'_B})].
\end{align}
\end{proposition}
\small
\textcolor{bblue}{Proof.}
\begin{align}
\frac{\kl(P_L\Vert P_{L'})}{\expop[w(L)]}\overset{(a)}{=}&\expop[\log_2\frac{Q_{S_B}Q'_B}{Q_BQ'_{S_B}}]\\
\overset{(b)}{=}&\expop[\log_2\frac{P_{S_B}(S_B)}{P_{S'_B}(S_B)}]\\
=&\expop[\kl(P_{S_B}\Vert P_{S'_B})]
\end{align}
where (a) follows by the differential LANSIT and (b) by \eqref{eq:qp}.\hfill$\square$
\end{frame}

\begin{frame}[plain]
\begin{center}
\color{bblue}{\Large Pinsker's Inequality for Trees}
\end{center}
\end{frame}

\begin{frame}
\frametitle{Variational distance}
$P_X$, $P_Y$ two distributions on $\mathcal{X}$. Variational distance $d(P_X,P_Y)$ is
\begin{align}
d(P_X,P_Y):=\sum_{a\in\mathcal{X}}|P_X(a)-P_Y(a)|
\end{align}
\textbf{Bounds:}
\begin{align}
&d(P_X,P_Y)\geq 0,\text{ with equality iff }\forall a\in\mathcal{X}\colon P_X(a)=P_Y(a)\\
&d(P_X,P_Y)\leq 2,\text{ with equality iff }\supp P_X\cap \supp P_Y=\emptyset.
\end{align}
\end{frame}

\begin{frame}
\frametitle{Approximating Distributions}
Set of distributions over $\mathcal{X}$: $\mathcal{P}_\mathcal{X}$.
\begin{proposition}
\begin{enumerate}[i]
\item Pinsker's Inequality:
\begin{align}
\kl(P_X\Vert P_Y)\geq \frac{1}{2\ln 2}d^2(P_X,P_Y).
\end{align}
\item Let $\{P_{X_k}\}_{k=1}^\infty$ be a set of distributions in $\mathcal{P}_\mathcal{X}$.
\begin{align}
\kl(P_{X_k}\Vert P_Y)\overset{k\to\infty}{\longrightarrow}0\quad&\Rightarrow\quad d(P_{X_k},P_Y)\overset{k\to\infty}{\longrightarrow}0
\end{align}
\item Let $g$ be a function on $\mathcal{P}_\mathcal{X}$ that is continuous in $P_Y$.
\begin{align}
\kl(P_{X_k}\Vert P_Y)\overset{k\to\infty}{\longrightarrow}0
\quad\Rightarrow\quad
\Bigl|g(P_{X_k})-g(P_Y)\Bigr|\overset{k\to\infty}{\longrightarrow}0.
\end{align}
\end{enumerate}
\end{proposition}
\end{frame}

\begin{frame}
\frametitle{Example: Entropy}
By \cite[Lemma 2.7]{csiszar2011information}, entropy is continuous in any distribution $P_Y\in\mathcal{P}_\mathcal{X}$. Thus
\begin{align}
\kl(P_{X_k}\Vert P_Y)\overset{k\to\infty}{\longrightarrow}0
\quad\Rightarrow\quad
|\entop(P_{X_k})-\entop(P_Y)|\overset{k\to\infty}{\longrightarrow}0.
\end{align}
\end{frame}

\begin{frame}
\frametitle{Product Distributions}
\begin{itemize}
\item Consider a tree and let $P_{S^*}$ be a branching distribution. Assign\footnote{This is a slight abuse of notation, since for $j\neq i$, $\mathcal{S}_i\neq \mathcal{S}_i$. However, we can think of $P_{S_j}$ as a distribution over branch labels. For example, for a binary tree, $P_{S_j}$ is then a distribution over the labels $\{0,1\}$, for all $i\in\mathcal{B}$ and the assignment $P_{S_j}=P_{S^*}$ is meaningful.} $P_{S_j}=P_{S^*}$ for all branching nodes $j\in\mathcal{B}$. We call the resulting node probabilities the \textcolor{bblue}{product distribution $P_{S^*}^+$}.
\item For any complete tree with leaves $\mathcal{L}$, $P_{S^*}^+$ defines a leaf distribution, i.e., $\sum_{i\in\mathcal{L}}P_{S^*}^+(i)=1$.
\item For any (possibly non-complete) tree with leaves $\mathcal{L}$, we define the informational divergence between the leaf distribution $P_L$ and $P_{S^*}^+$ as
\begin{align}
\kl(P_L\Vert P_{S^*}^+):=\sum_{i\in\mathcal{L}}P_L(i)\log_2\frac{P_L(i)}{P_{S^*}^+(i)}.
\end{align}
\end{itemize}
\end{frame}
\begin{frame}
\frametitle{Approximating Distributions on Trees}
\begin{proposition}\label{prop:pinskertree}
\begin{enumerate}[i]
\item Pinsker's Inequality for Trees:
\begin{align}
\frac{\kl(P_L\Vert P_{L'})}{\expop[w(L)]}\geq\frac{1}{2\ln(2)}\expop[d^2(P_{S_B},P_{S'_B})].
\end{align}
\item For any $\epsilon>0$,
\begin{align}
\frac{\kl(P_L\Vert P_{L'})}{\expop[w(L)]}\overset{|\mathcal{L}|\to\infty}{\longrightarrow}0\;\Rightarrow\;\probop[d(P_{S_B},P_{S'_B})\geq\epsilon]\overset{|\mathcal{L}|\to\infty}{\longrightarrow}0.
\end{align}
\item Let $P_{S^*}$ be a branching distribution and let $g$ be a function on $\mathcal{P}_\mathcal{S}$ that is bounded and continuous in $P_{S^*}$.
\begin{align}
\frac{\kl(P_L\Vert P_{S^*}^+)}{\expop[w(L)]}\overset{|\mathcal{L}|\to\infty}{\longrightarrow}0
\;\Rightarrow\;
\Bigl|\expop[g(P_{S_B})]-g(P_{S^*})\Bigr|\overset{|\mathcal{L}|\to\infty}{\longrightarrow}0.
\end{align}
\end{enumerate}
\end{proposition}
\textcolor{bblue}{Proof.} See Slides \ref{slide:prooffirst}--\ref{slide:prooflast}. \hfill$\square$
\end{frame}

\begin{frame}
\frametitle{Entropy Rate}
By Prop.~\ref{prop:dlel},
\begin{align}
\frac{\entop(P_L)}{\expop[w(L)]}&=\expop[\entop(P_{S_B})].
\end{align}
$\entop$ is continuous and bounded. Thus by Prop.~\ref{prop:pinskertree}iii. we have the following proposition.
\begin{proposition}\label{prop:entropyrateconvergence}
\begin{align}
\frac{\kl(P_L\Vert P_{S^*}^+)}{\expop[w(L)]}\overset{|\mathcal{L}|\to\infty}{\longrightarrow}0\quad\Rightarrow\quad\Bigl|\frac{\entop(P_L)}{\expop[w(L)]}-\entop(P_{S^*})\Bigr|\overset{|\mathcal{L}|\to\infty}{\longrightarrow}0.
\end{align}
\end{proposition}
\end{frame}

\begin{frame}
\frametitle{Random Vectors}
\textcolor{bblue}{Remark.} (See also Slide \ref{slide:remark}) If all paths in a tree have the same length $n$,  then $P_L$ can be thought of as a joint distribution $P_{X^n}$ of a random vector $X^n=(X_1,\dotsc,X_n)$. The (tree) product distribution $P_{S^*}^+$ is then the conventional product distribution $P_{S^*}^n$. Prop.~\ref{prop:pinskertree} applies and in particular, Prop.~\ref{prop:entropyrateconvergence} becomes
\begin{align}
\frac{\kl(P_{X^n}\Vert P_{S^*}^n)}{n}\overset{n\to\infty}{\longrightarrow}0\quad\Rightarrow\quad\Bigl|\frac{\entop(P_{X^n})}{n}-\entop(P_{S^*})\Bigr|\overset{n\to\infty}{\longrightarrow}0.
\end{align}
\end{frame}

\begin{frame}
\frametitle{Proof of Prop.~\ref{prop:pinskertree}i.}
\label{slide:prooffirst}
\begin{align}
\frac{\kl(P_L\Vert P_{L'})}{\expop[w(L)]} \overset{(a)}{=} &\expop[\kl(P_{S_B}\Vert P_{S'_B})]\\
\overset{(b)}{\geq}&\frac{1}{2\ln 2}\expop[d^2(P_{S_B},P_{S'_B})]
\end{align}
where (a) follows by Prop.~\ref{prop:ndivergence} and where (b) follows by Pinsker's inequality.\hfill$\square$
\end{frame}

\begin{frame}
\frametitle{Proof of Prop.~\ref{prop:pinskertree}ii.}
\label{slide:proofii}
Suppose $\expop[d(P_{S_B},P_{S'_B})]<\epsilon^2$ for some $\epsilon>0$. Then
\begin{align}
\probop[d(P_{S_B},P_{S'_B})\geq\epsilon]&\overset{(a)}{\leq}\frac{\expop[d(P_{S_B},P_{S'_B})]}{\epsilon}\\
&\leq\frac{\epsilon^2}{\epsilon}\\
&=\epsilon
\end{align}
where (a) follows by Markov's inequality \cite[Theo.~A.2]{kramer2012information}. Together with statement i., statement ii. follows.\hfill$\square$
\end{frame}

\begin{frame}
\frametitle{Proof of Prop.~\ref{prop:pinskertree}iii. (1)}
By assumption, $g$ is bounded and continuous in $P_{S^*}$. By boundedness, there exists a value $g_{\max}<\infty$ such that
\begin{align}
\forall j\in\mathcal{B}\colon|g(P_{S_j})-g(P_{S^*})|&\leq g_{\max}.\label{eq:iii:bounded}
\end{align}
By continuity, we know that
\begin{align}
\forall \delta>0\colon\exists \epsilon_\delta\colon&\forall \epsilon'<\epsilon_\delta\colon\nonumber\\ &d(P_{S_j},P_{S^*})<\epsilon'\Rightarrow |g(P_{S_j})-g(P_{S^*})|<\delta.\label{eq:iii:cont}
\end{align}
Define 
\begin{align}
\epsilon=\min\{\epsilon_\delta,\delta\}.\label{eq:iii:epsmin}
\end{align}
\end{frame}

\begin{frame}
\frametitle{Proof of Prop.~\ref{prop:pinskertree}iii. (2)}
Suppose $\expop[d(P_{S_B},P_{S^*})]<\epsilon^2$. We write
\begin{align} 	
&\hspace{-0.5cm}|\expop[g(P_{S_B})]-g(P_{S^*})|=\Bigl|\sum_{j\in\mathcal{B}} P_B(j)[g(P_{S_j})-g(P_{S^*})]\Bigr|\nonumber\\
\leq&\sum_{j\in\mathcal{B}} P_B(j)\bigl|g(P_{S_j})-g(P_{S^*})\bigr|\\
=&\sum_{j\colon d(P_{S_j},P_{S^*})<\epsilon}P_B(j)\bigl|g(P_{S_j})-g(P_{S^*})\bigr|\nonumber\\
&\qquad+\sum_{j\colon d(P_{S_j},P_{S^*})\geq\epsilon}P_B(j)\bigl|g(P_{S_j})-g(P_{S^*})\bigr|.\label{eq:iii:sums}
\end{align}
We next bound the two sums in \eqref{eq:iii:sums}.
\end{frame}

\begin{frame}
\frametitle{Proof of Prop.~\ref{prop:pinskertree}iii. (3)}
The first sum in \eqref{eq:iii:sums} is bounded as
\begin{align} 	
\sum_{j\colon d(P_{S_j},P_{S^*})<\epsilon}\!\!P_B(j)\bigl|g(P_{S_j})-g(P_{S^*})\bigr|\overset{(a)}{\leq} &\sum_{j\colon d(P_{S_j},P_{S^*})<\epsilon}\!\!P_B(j) \delta\nonumber\\
\leq &\delta\label{eq:iii:stsum}
\end{align}
where (a) follows by \eqref{eq:iii:cont} and \eqref{eq:iii:epsmin}.
\end{frame}

\begin{frame}
\frametitle{Proof of Prop.~\ref{prop:pinskertree}iii. (4)}
The second sum in \eqref{eq:iii:sums} is bounded as
\begin{align} 	
\sum_{j\colon d(P_{S_j},P_{S^*})\geq\epsilon}P_B(j)\bigl|g(P_{S_j})-g(P_{S^*})\bigr|\overset{(a)}{\leq} &\sum_{j\colon d(P_{S_j},P_{S^*})\geq\epsilon}P_B(j) g_{\max}\nonumber\\
\overset{(b)}{\leq} &\epsilon g_{\max}\nonumber\\
\overset{(c)}{\leq} &\delta g_{\max}\label{eq:iii:ndsum}
\end{align}
where (a) follows by \eqref{eq:iii:bounded}, where (b) follows by our assumption $\expop[d(P_{S_B},P_{S^*})]<\epsilon^2$ and Slide~\ref{slide:proofii} and where (c) follows by \eqref{eq:iii:epsmin}.
\end{frame}

\begin{frame}
\frametitle{Proof of Prop.~\ref{prop:pinskertree}iii. (5)}
\label{slide:prooflast}
Using \eqref{eq:iii:stsum} and \eqref{eq:iii:ndsum} in \eqref{eq:iii:sums}, we get
\begin{align}
|\expop[g(P_{S_B})]-g(P_{S^*})|\leq \delta + \delta g_{\max}=\delta(1+g_{\max}).
\end{align}
For $\delta\to 0$, the error bound on the right-hand side goes to zero, which proves part iii. of Prop.~\ref{prop:pinskertree}.\hfill$\square$
\end{frame}

\begin{frame}
\frametitle{References}
\bibliographystyle{IEEEtran}
\normalsize
\bibliography{IEEEabrv,confs-jrnls,references}

\begin{thebibliography}{1}
\providecommand{\url}[1]{#1}
\csname url@samestyle\endcsname
\providecommand{\newblock}{\relax}
\providecommand{\bibinfo}[2]{#2}
\providecommand{\BIBentrySTDinterwordspacing}{\spaceskip=0pt\relax}
\providecommand{\BIBentryALTinterwordstretchfactor}{4}
\providecommand{\BIBentryALTinterwordspacing}{\spaceskip=\fontdimen2\font plus
\BIBentryALTinterwordstretchfactor\fontdimen3\font minus
  \fontdimen4\font\relax}
\providecommand{\BIBforeignlanguage}[2]{{%
\expandafter\ifx\csname l@#1\endcsname\relax
\typeout{** WARNING: IEEEtran.bst: No hyphenation pattern has been}%
\typeout{** loaded for the language `#1'. Using the pattern for}%
\typeout{** the default language instead.}%
\else
\language=\csname l@#1\endcsname
\fi
#2}}
\providecommand{\BIBdecl}{\relax}
\BIBdecl

\bibitem{rueppel1994leaf}
R.~A. Rueppel and J.~L. Massey, ``Leaf-average node-sum interchanges in rooted
  trees with applications,'' in \emph{Communications and Cryptography: Two
  sides of One Tapestry}, R.~E. Blahut, D.~J. {Costello Jr.}, U.~Maurer, and
  T.~Mittelholzer, Eds.\hskip 1em plus 0.5em minus 0.4em\relax Kluwer Academic
  Publishers, 1994.

\bibitem{masseyapplied1}
\BIBentryALTinterwordspacing
J.~L. Massey, ``Applied digital information theory {I},'' lecture notes, {ETH
  Zurich}. [Online]. Available:
  \url{http://www.isiweb.ee.ethz.ch/archive/massey_scr/adit1.pdf}
\BIBentrySTDinterwordspacing

\bibitem{kramer2012information}
G.~Kramer, ``Information theory,'' lecture notes TU Munich, edition WS
  2012/2013.

\bibitem{csiszar2011information}
I.~Csisz\'ar and J.~K\"orner, \emph{Information Theory: Coding Theorems for
  Discrete Memoryless Systems}.\hskip 1em plus 0.5em minus 0.4em\relax
  Cambridge University Press, 2011.

\end{thebibliography}
\end{frame}

\end{document}